\documentclass[
    ,final            % use final for the camera ready runs
  ]
  {aipproc}

\layoutstyle{6x9}

\begin{document}

\title{Jefferson Lab's results on the $Q^2$-evolution of moments of spin
structure functions}


\author{A. Deur}{
  address={Thomas Jefferson National Accelerator Facility, 
Newport-News, VA 23606, USA}
  }

\begin{abstract}
We present the recent JLab measurements on moments of spin structure functions 
at intermediate and low $Q^2$. The Bjorken sum and Burkhardt-Cottingham sum on
the neutron are presented. The later appears to hold. Higher moments 
(generalized spin polarizabilities and $d_2^n$) are shown and compared to 
chiral perturbation theory and lattice QCD respectively. 
\end{abstract}

\maketitle

%%%%%%%%%%%%%%%%%%%%%%%%%%%%%%%%%%%%%%%%%%%%
%% MAINMATTER
%%%%%%%%%%%%%%%%%%%%%%%%%%%%%%%%%%%%%%%%%%%%
\vspace{-1 cm}
\section{Moments of Spin Structure Functions}
\vspace{-0.3 cm}
Polarized DIS has provided a testing ground for the study of the strong force. 
Moments of spin structure functions (SSF), among them the Bjorken sum, has 
played an important r\^ole in this study. The n-th (Cornwall-Norton) moment of 
SSF is the integral of the $x^n g_{1,2}(x,Q^2)$ SSF over $x$. Moments are 
specially useful because sum rules relate them to other quantities. Such sum 
rules for $\Gamma_1$, the first moment of $g_1$, are the 
Ellis-Jaffe~\cite{EJSR} and the Bjorken sum rules~\cite{Bjorken}, derived at 
large $Q^2$, and the related Gerasimov-Drell-Hearn (GDH) sum rule~\cite{GDH} 
at $Q^2=0$. The first moment of $g_2$, $\Gamma_2$, is given by the 
Burkhardt-Cottingham (BC) sum rule~\cite{BC}. Rules can be also derived for 
higher moments, e.g., spin polarizability or $d_2$ sum rules. \\
These relations are useful in many ways: checks the theory on which the rule 
is based (e.g. QCD and the Bjorken sum rule); checks hypotheses used in the 
sum rule derivation (e.g. the Ellis-Jaffe sum rules); or checks calculations 
such as chiral perturbation theory ($\chi pt$), lattice QCD or Operator Product 
Expansion (OPE). If a sum rule rests on solid grounds or is well tested, it 
can be used to extract quantities otherwise hard to measure (e.g. generalized 
spin polarizabilities). Because $\Gamma_{1,2}$ are calculable at any $Q^2$ 
using either $\chi pt$, lattice QCD or OPE, they are particularly suited to 
study the transition between the hadronic to partonic descriptions of the 
strong force. Measurements in the transition region (intermediate $Q^2$) 
have recently been made at Jefferson Lab (JLab).
\section{Measurements at Jefferson Lab}
\vspace{-0.3 cm}
At moderate $Q^2$, resonances saturate moments. JLab's accelerator 
delivers CW electron beam with a maximum energy up to 6 GeV. This makes JLab 
the suited place to measure moments up to $Q^2$ of a few GeV$^2$. The beam 
current can reach 200 $\mu$A with a polarization now reaching 85\% although at 
the time of the experiments reported here, it was typically 70\%. 
The beam is sent simultaneously to three halls (A, B and C), all of them 
equipped with polarized targets. In this talk, we report on results from halls 
A and B.\\
Hall A~\cite{HallA nim} contains a polarized $^3$He gaseous target and two 
high resolution spectrometers (HRS) with 6 mSr acceptance. The target can be 
polarized longitudinally or transversally at typically 40\% polarization with 
10-15  $\mu$A of beam. The target's $\sim10$ atm. of $^3$He gives a 
luminosity greater than $10^{36}$cm$^{-2}$s$^{-1}$. Hall B~\cite{HallB nim} 
luminosity is typically 5$\times 10^{33}$cm$^{-2}$s$^{-1}$ but is compensated 
by the large acceptance (about 2.5$\pi$) of the CLAS spectrometer. Cryogenic 
polarized targets (NH$_3$ and  ND$_3$) are well suited for the low beam 
currents ($\sim$nA) utilized in Hall B. The target is longitudinally polarized 
with average 75\% (NH$_3$) and 40\% (ND$_3$) polarizations. Both halls can 
cover the large region of $Q^2$ and $x$ needed to extract moments at various 
$Q^2$, either because of the large CLAS acceptance (Hall B) or because of large 
luminosity allowing to quickly gather data at various beam energies 
and HRS settings (Hall A).\\
I report here on the Hall A E94010~\cite{E94010} and Hall B EG1 experiments. 
EG1 was split in two runs: EG1a (1998) which results are 
published~\cite{eg1a}, and EG1b (2000) that is still being analyzed. SSF are 
extracted differently in halls A and B. In Hall A, \emph{absolute} cross 
sections asymmetries $\Delta \sigma^{\| (\bot)}$ were measured for 
longitudinal (transverse) target spin orientations. $g_1$ and $g_2$ are linear 
combinations of these $\Delta \sigma$ and are extracted without external 
input. Furthermore, unpolarized contributions, e.g. target cell windows or the 
(mostly) unpolarized protons in the $^3$He nucleus, cancel out. The 
\emph{relative} longitudinal asymmetry $A_{\|}$ is measured in Hall B. Models 
for $F_1$, $g_2$ and $R=\sigma _L / \sigma _T$ are then used to extract 
$g_1$. $F_1$ and $R$ are constrained at low $Q^2$ by recent Hall C 
data~\cite{E94110}. $g_2$ is estimated using models (resonance region) or its 
leading twist part $g_2^{ww}$ (DIS domain). The unmeasured low-$x$ part of the 
moment is estimated using a parametrization developed by the EG1 
collaboration, while the E94010 group used a Regge-type fit of DIS 
data~\cite{Bianchi}.
\begin{figure}
\includegraphics[height=.48\textheight]{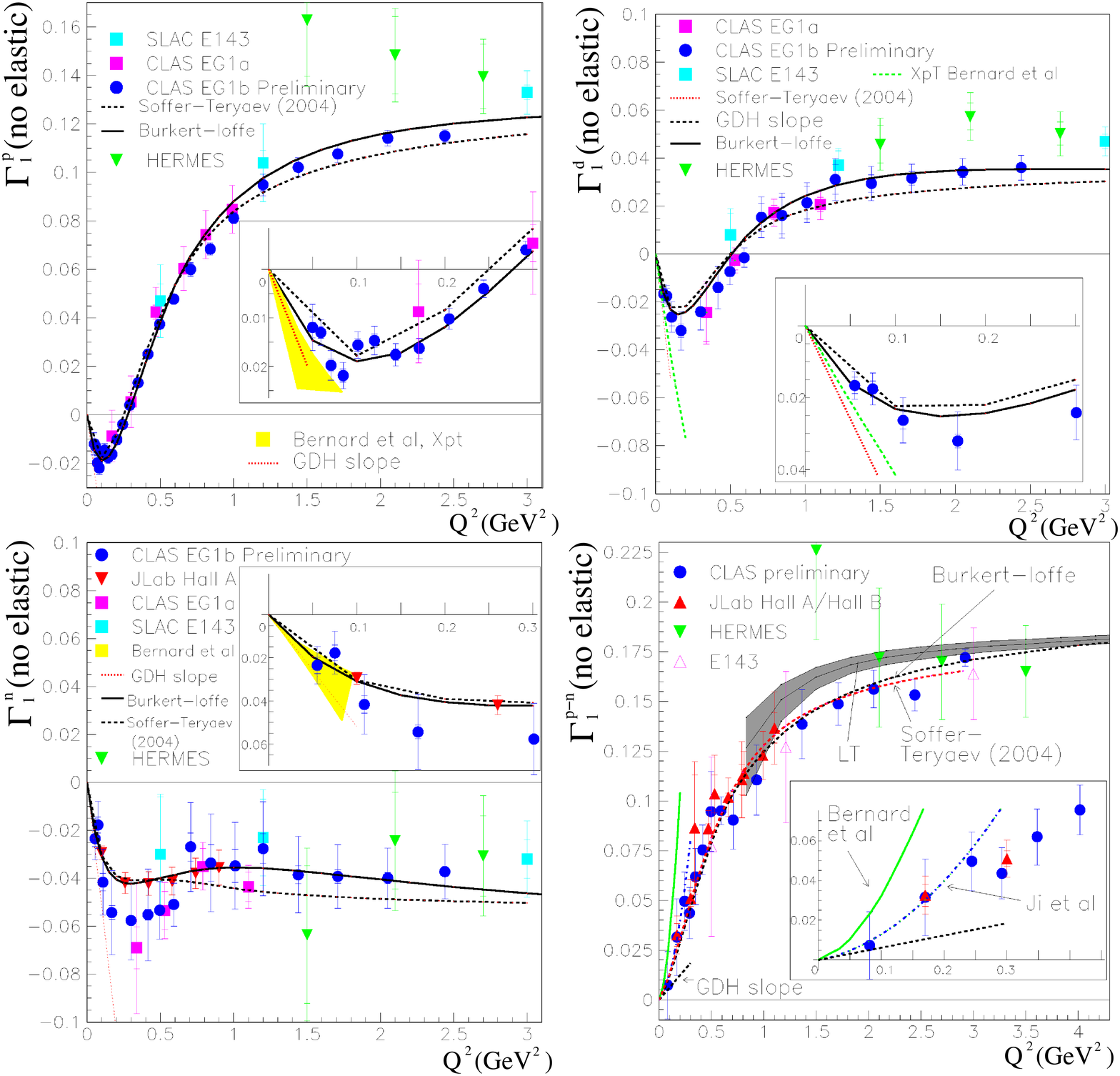}
\caption{First moments $\Gamma_1^{p}$, $\Gamma_1^{n}$, $\Gamma_1^{d}$ and 
the Bjorken sum $\Gamma_1^{p-n}$. The elastic contribution is excluded.}
\end{figure}
Results on $\Gamma_1^{p}$, $\Gamma_1^{n}$ and $\Gamma_1^{d}$ are shown in 
Fig. 1, together with $\chi pt$ calculations~\cite{meissner chipt,ji chipt} 
models~\cite{Burkert and Ioffe,Soffer} and leading twist OPE prediction. 
HERMES~\cite{HERMES} and SLAC~\cite{e143} results are also shown. The halls A 
and B data, reanalyzed at matched $Q^2$ points and with a consistent low-$x$ 
estimate~\cite{Bianchi} were used to form the Bjorken sum 
$\Gamma_1^{p-n}$~\cite{deur}. Preliminary $\Gamma_1^{p-n}$ from EG1b is also 
shown. $\Gamma_1^{p-n}$ is a unique quantity to study parton-hadron transition 
because its non-singlet structure makes it an easier quantity to handle for 
$\chi pt$, lattice QCD and OPE. These data form, for both nucleons, an 
accurate mapping at intermediate $Q^2$  that connects to SLAC, HERMES and CERN 
DIS data. At low $Q^2$, $\chi pt$ disagrees with the data above $Q^2=0.2$ 
GeV$^2$, while models based on different physics reproduce equally well the 
data. Twist 2 description also works well down to low $Q^2$, indicating an 
overall suppressed higher twist r\^ole. Indeed, in OPE analysis 
results~\cite{deur,osipenko,ZEM}, twist 4 and 6 coefficients are either small 
or canceling each others at $Q^2$=1 GeV$^2$.\\
The availability of transverse data in Hall A allows us to form $\Gamma_2^n$ 
and thereby check the BC sum rule $(\Gamma_2=0)$ on the neutron (fig. 2). The 
sum rule is based on dispersion relations and is $Q^2$-invariant. A striking 
feature is the almost perfect cancellation between elastic and resonance 
contributions leading to the verification of the sum rule. 
\begin{figure}
\includegraphics[height=.29\textheight]{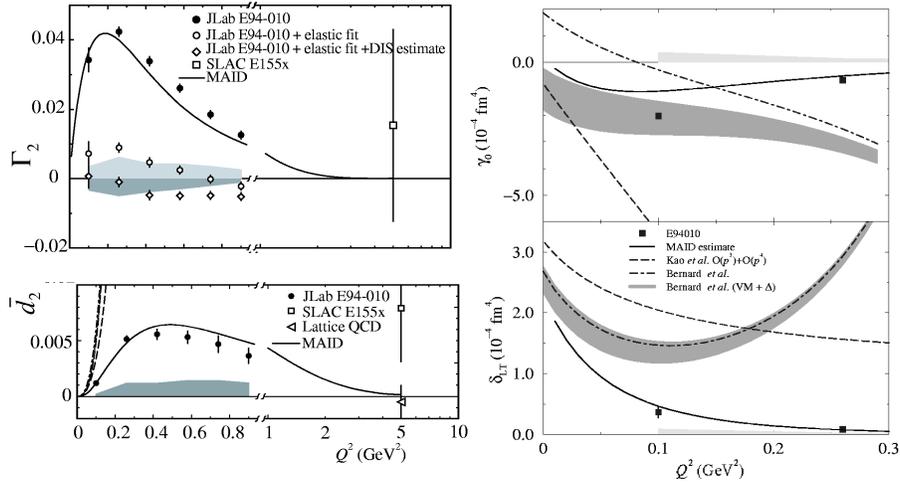}
\caption{Moments $\Gamma_2^{n}$ and $\overline{d}_2^n$ (left), and 
generalized spin polarizabilities $\gamma_0$ and $\delta_{LT}$ (right)}
\end{figure}
Other sum rules link SSF moments to the generalized spin polarizabilities 
$\gamma_0$ and $\delta_{LT}$:
\begin{eqnarray}
\gamma_{0}=\frac{4e^{2}M^{2}}{\pi Q^{6}}
\int_{0}^{1^-}x^{2}(g_{1}-\frac{4M^{2}}{Q^{2}}x^2g_{2})dx~~;~~
\delta_{LT}=\frac{4e^{2}M^{2}}{\pi Q^{6}}
\int_{0}^{1^-}x^{2}(g_{1}+g_{2})dx
\nonumber
\end{eqnarray}
Results from Hall A can be seen in fig. 2~\cite{E94010-3}. $\delta_{LT}$ is 
interesting because the $\Delta_{1232}$ r\^ole is suppressed. Hence 
$\delta_{LT}$ is easier to access by $\chi pt$. However, calculations and data 
disagree for both $\gamma_0$ and $\delta_{LT}$. The MAID model~\cite{MAID}, 
however, well reproduces the data. Another higher moment that can be formed is 
$d_2^n$, the integral of $x^{2}(g_{2} - g_{2}^{ww})$ where $g_2^{ww}$ is the 
leading twist part of $g_2$. Thus $d_2$ is sensitive to twist 3 and higher. 
The measured $\overline{d}_2^n$ (the bar indicates the exclusion of $x=1$) 
trends toward the lattice QCD results, although larger $Q^2$ data are 
necessary to establish a possible agreement.
\section{Summary and Perspectives}
\vspace{-0.3 cm}
The hadron-parton transition region is covered by data of the SSF moments from 
JLab. These can be calculated at any $Q^2$, thus providing a ground for 
studying the link between hadronic and partonic descriptions of the strong 
force. An OPE analysis reveals that in this domain, high twist effects are 
small. The BC sum rule was shown on the neutron and found to be valid. Data 
and sum rules were used to extract neutron generalized spin polarizabilities. 
Those disagree with the present $\chi pt$ calculations. Further data from Hall 
A E01-012~\cite{e01012}, Hall B EG1b, and Hall C RSS~\cite{RSS} will be 
available shortly in the resonance region. New data at very low $Q^2$ have been 
taken on the neutron in Hall A~\cite{e97110} and will be gathered early 2006 
for the proton in Hall B~\cite{e03006}. The 12 GeV upgrade of JLab will allow 
us to access both larger-$x$ and lower-$x$. This will allow for more precise 
measurements of the moments, in particular by addressing the low$x$ issue.  
\begin{theacknowledgments}
\vspace{-0.3 cm}
\footnotesize{
This work is supported by the U.S. Department of Energy (DOE) and the U.S.
National Science Foundation. The Southeastern Universities Research 
Association operates the Thomas Jefferson National Accelerator 
Facility for the DOE under contract DE-AC05-84ER40150.} 
\end{theacknowledgments}
\bibliographystyle{aipproc}   % if natbib is available

%%%%%%%%%%%%%%%%%%%%%%%%%%%%%%%%%%%%%%%%%%%
%% The following lines show an example how to produce a bibliography
%% without the help of the BibTeX program. This could be used instead
%% of the above.
%%%%%%%%%%%%%%%%%%%%%%%%%%%%%%%%%%%%%%%%%%%

\end{document}